\documentclass{desyproc}

\begin{document}
\title{\vspace{-2.05cm}
\hfill{\small{DESY 12-234}}\\[1.27cm]
BSM photon interaction for ALPS-II and beyond}

\author{{\slshape Babette D\"obrich$^1$}\\[1ex]
$^1$Deutsches Elektronen-Synchrotron (DESY), Notkestr. 85, 22607 Hamburg, Germany}

\contribID{doebrich\_babette}

\desyproc{DESY-PROC-2012-04}
\acronym{Patras 2012} 
\doi  

\maketitle

\begin{abstract}
High-intensity photon beams can provide for a viable probe for many particles of 
Standard Model extensions. 
This workshop contribution briefly reviews the status of the second stage of the
Any Light Particle Search (ALPS-II) at DESY, 
an experiment of the light-shining-through-a-wall 
type, as well as an idea to test asymptotically safe quantum gravity in a 
photon-scattering experiment.
\end{abstract}

\section{An enlightening way of looking for new physics}
Several considerations (e.g., UV-completions for the Standard Model) and 
observations (e.g., Dark Matter)
lead us to assume the existence of particles beyond the Standard Model.
In order to have evaded detection so far, such new particles can 
be either very heavy, or rather
light (e.g., at sub-eV scale) if they have
extremely small coupling to known particles \cite{Jaeckel:2010ni}. 

The most renowned example 
of such proposed weakly interacting slim particles, ``WISPs'' for short, 
is arguably the axion \cite{Wilczek:1977pj}, 
which is a consequence of the Peccei-Quinn solution to the strong CP-problem.
In addition, strong interest has emerged recently for so-called 
axion-{\it like} particles (ALPs), whose mass-coupling 
relation is relaxed compared to
the QCD axion: 
ALPs can, e.g., appear in intermediate string 
scale scenarios \cite{Cicoli:2012sz},
and
constitute Dark Matter \cite{Arias:2012az}. Moreover they could 
explain yet puzzling observations\footnote{Note that
astrophysical observations
of course also put strong constraints on the existence of ALPs.
A recent comprehensive overview of the corresponding
parameter space can be found in \cite{Hewett:2012ns}.}  
in some astrophysical processes
such as anomalous White Dwarf cooling \cite{Isern:2008nt} and the 
transmissibility of the universe to 
high-energetic photons \cite{Horns:2012fx}.

Further WISPs can be particles of hidden sectors in string- and field-theoretic 
extensions of the 
Standard Model, see, e.g., \cite{Abel:2008ai}: 
Particularly hidden photons (HPs), i.e., gauge bosons 
of an extra U(1) gauge group
as well hidden sector matter. The latter
can acquire an electromagnetic fractional charge
and thus can constitute so-called minicharged particles (MCPs), 
see  \cite{Jaeckel:2010ni}
for an overview. Hidden photons are also a 
viable Dark Matter candidate \cite{Arias:2012az}
and could be responsible for the phenomenon of 
Dark Radiation \cite{Jaeckel:2008fi}.
In addition, WISPs can appear as scalar modes in theories of massive gravity \cite{Brax:2012ie}.

If such WISPs exist, it is expedient to search for 
them also by their interactions 
with photons. Amongst others, this is advantageous 
because photons can be easily produced  at high rates and do not have tree-level self-interactions within the Standard Model.
Thus, beyond-Standard-Model physics becomes readily accessible.
Experiments particularly apt to look for WISPs with photons are of the 
`light-shining-through-a-wall'-type (LSW) \cite{Redondo:2010dp}:
Laser photons can be converted into a WISP in front of a light-blocking 
barrier (generation region)
and reconverted into photons behind that barrier (regeneration region). 
Depending on the particle type, these
conversion processes are induced by magnetic 
fields\footnote{Note that
for the test of ALPs and MCPs the optimal direction of these fields 
is rather different \cite{Dobrich:2012sw}.} or are manifest as oscillations.

The most sensitive LSW laboratory setup thus far is the first 
stage of the Any Light Particle Search (ALPS-I) \cite{Ehret:2010mh} at DESY. 
With major upgrades in magnetic length, laser power and the detection system, 
the proposed ALPS-II experiment
aims at improving the sensitivity by a few orders of magnitude for the 
different WISPs.
Following last year's workshop contribution  \cite{vonSeggern:2011em} we shortly 
present an updated status of ALPS-II.

\section{ALPS-II status and prospect}

The reason for proposing to realize an upgraded version of 
ALPS is its sensitivity 
to particularly interesting parameter regions for various WISPs, 
as indicated in the previous section.

Three key ingredients are responsible for this sensitivity boost \cite{TDR}, 
cf. Tab.~\ref{tab:param}: 
Foremost, the magnetic length of
ALPS-II is expected to be 468~Tm. This can be achieved by a string of 
10+10 HERA dipoles which can be taken 
from a reserve of 24 spare magnets manufactured for HERA at DESY.
Note that the setup of this string requires an aperture increase by
 straightening the beam pipe
to avoid clipping losses for the laser. 
The viability of this undertaking has been proven with the 
straightened ALPS-I magnet
achieving quench currents above values measured for the unbent magnet  \cite{TDR}.
Secondly, the effective photon flux in the setup is planned  \cite{baehre} to be increased 
through a higher power buildup in the production cavity and
by virtue of `resonant regeneration' \cite{Hoogeveen:1990vq}, 
i.e., an optical resonator in the 
regeneration region,
locked to the resonator in the 
generation region\footnote{Note that `resonant regeneration' is already 
successfully used in 
related setups with microwaves \cite{Betz:2012tp}. For the optical regime,
different locking schemes have been proposed \cite{TDR,Mueller:2010zzc}.}.
To assure long-term stability of the cavity mirrors, 
ALPS-II will employ infrared light at 1064nm wavelength
(instead of green 532nm for ALPS-I).
Thirdly, ALPS-II will feature a nearly background-free transition edge sensor
allowing for high detection efficiency even to infrared light. 
Whilst this sensor is under development,
the ALPS-I CCD can be used as a fall-back option.

\begin{table}
\begin{tabular}{|l|c|c|c|c|c|}
\hline
Parameter & Scaling & ALPS-I & ALPS-IIc & Sens. gain \\ [1pt] \hline
Effective laser power $P_{\rm laser}$ & $g_{a\gamma} \propto P_{\rm laser}^{-1/4}$ & 1\,kW & 150\,kW & 3.5\\[1pt] \hline
Rel. photon number flux $n_\gamma$ & $g_{a\gamma} \propto n_\gamma^{-1/4}$ & 1 (532\,nm) & 2 (1064\,nm) & 1.2\\[1pt] \hline
Power built up in RC $P_{\rm RC}$ & $g_{a\gamma} \propto P_{reg}^{-1/4}$ & 1 &  40,000 &  14\\[1pt] \hline
 $BL$ (before\& after the wall) & $g_{a\gamma} \propto (BL)^{-1}$ & 22\,Tm & 468\,Tm  & 21\\[1pt] \hline
Detector efficiency $QE$ & $g_{a\gamma} \propto QE^{-1/4}$ & 0.9 & 0.75 & 0.96\\[1pt] \hline
Detector noise $DC$ & $g_{a\gamma} \propto DC^{1/8}$ & 0.0018\,s$^{-1}$ & 0.000001\,s$^{-1}$ & 2.6\\[1pt] \hline
Combined improvements & &  &  & 3082\\[1pt] \hline
\end{tabular}
\caption{Parameters of the ALPS-I experiment in comparison to the ALPS-II proposal. 
The second column shows 
the dependence of the reachable ALP-photon coupling on the experimental parameters.
The last column lists the approximate 
sensitivity gain for ALP searches compared to ALPS-I. 
For hidden photons, there is no gain from the magnetic field. Thus the 
sensitivity gain follows as above except for the factor coming from $BL$
and amounts to 147.}
\label{tab:param}
\end{table}

\begin{figure*}[htbp]
\begin{tabular}{cc}
\begin{minipage}[t]{0.48\textwidth}
\begin{center}
\includegraphics[width=1\textwidth]{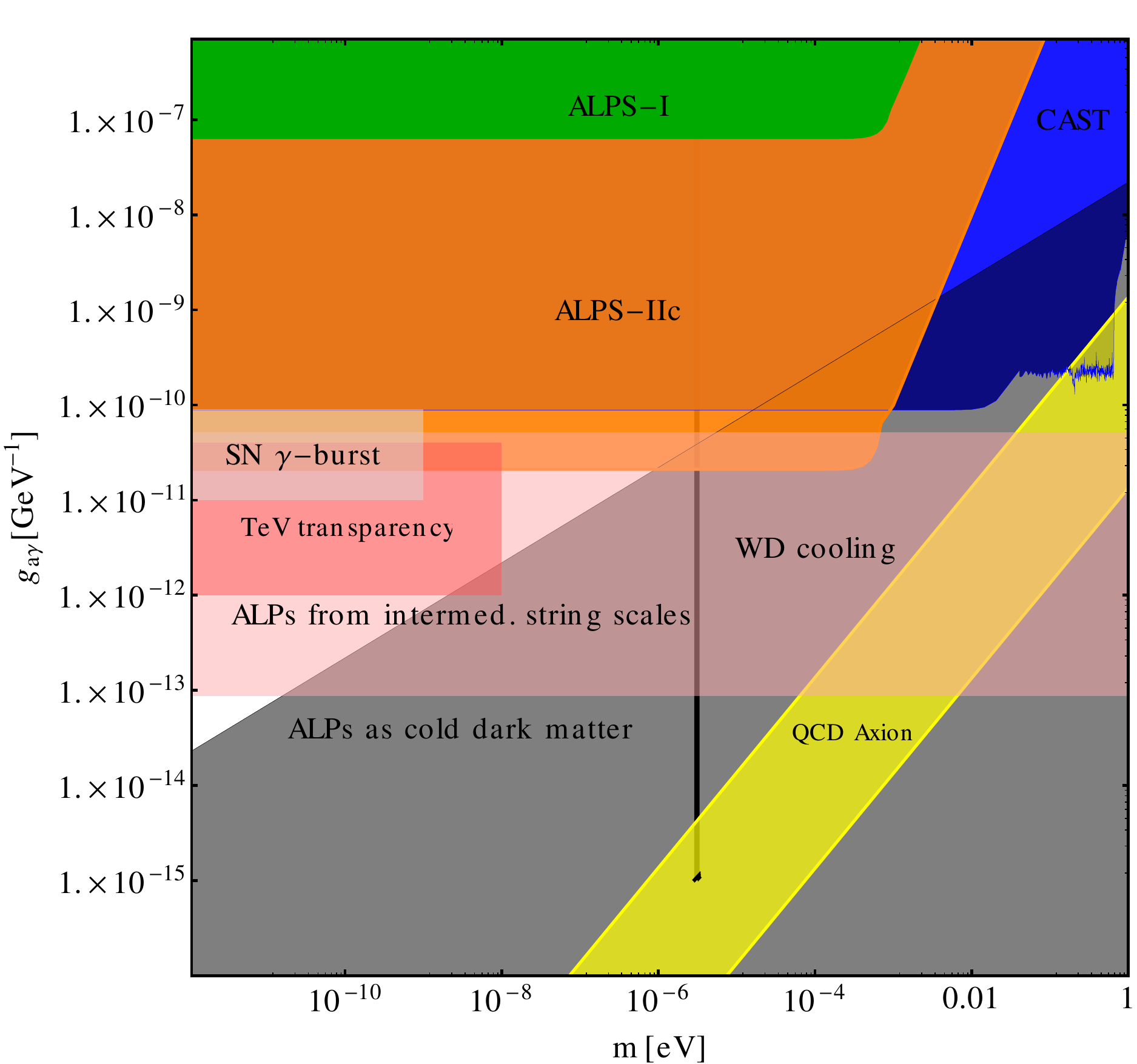}
\caption{Sketch of the prospective reach of ALPS-IIc (orange) in the axion-like particle parameter space,
see text for details.
} 
\label{fig:ALPs}
\end{center}
\end{minipage}

&

\begin{minipage}[t]{0.5\textwidth}
\begin{center}
\includegraphics[width=1\textwidth]{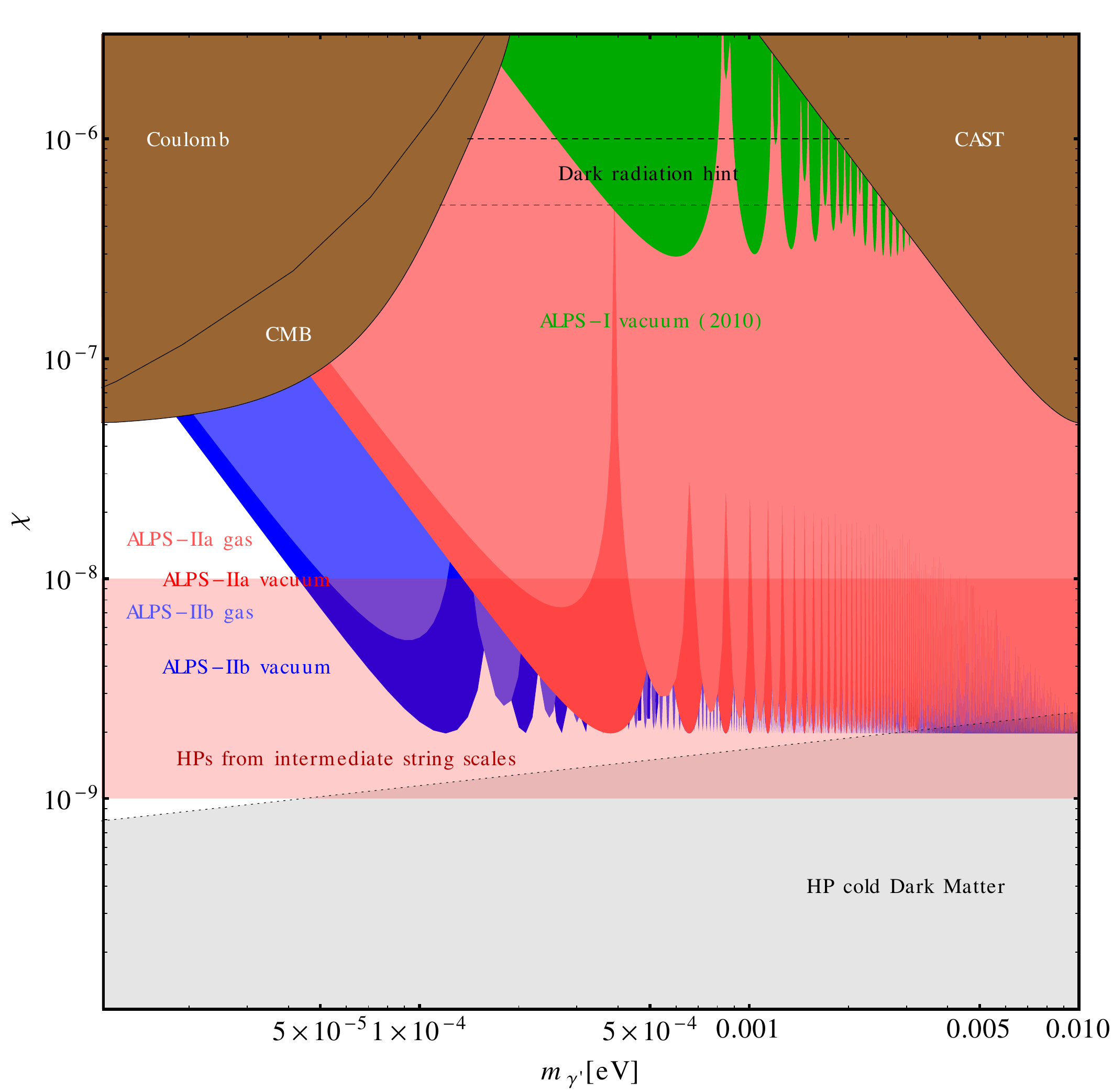} 
\caption{Expected sensitivity range of ALPS-IIa/b (red/blue) for hidden photons,
cf. text for details.
}
\label{fig:HPs}
\end{center}
\end{minipage}

\end{tabular}
\end{figure*}

ALPS-II is set out to be realized in three stages:
ALPS-IIa is already well under way and will search for hidden photons
in a 10m+10m LSW-setup without magnets. 
It is meant to demonstrate
the viability of the optics setup, particularly the locking scheme of the regeneration 
cavity.
As physics benefit, it will be sensitive to the Dark Radiation 
hint \cite{Jaeckel:2008fi} as well as
to parameter regions favored in intermediate string scale scenarios and a small region compatible with
HP Dark Matter, see Fig.~\ref{fig:HPs}.
As already successfully used at ALPS-I \cite{Ehret:2010mh}, also gas will be inserted into the vacuum
tube at each stage to close the gaps arising from 
at minima of the conversion probability.
ALPS-IIb, still without magnets is planned to test a smooth operation at 100m+100m length in the HERA tunnel.
Only the final stage, ALPS-IIc, is planned to operate with magnets and thus explore 
the parameter space of axion-like particles,
cf. Fig.~\ref{fig:ALPs}.

\section{Photons as probe for asymptotically safe gravity}

A further field of search for physics beyond the Standard Model
in which lasers-based setups could  
yield insight, is the search for a quantum theory of gravity.
As there is no tree-level background within the Standard Model,
it is tempting to explore if measuring the cross-section of photon-photon scattering at 
high energies can teach us about quantum gravity (QG), since the small QG signal is easier 
accessible without a tree-level background.

To achieve high photon energies in a collider mode,
different future options are advanced:
Compton-backscattering is possible within linacs or from wake-field-accelerated charged particles 
using pulsed high-intensity lasers \cite{Kando:2008te}.
The tiny cross section for photon-photon scattering through graviton exchange  \cite{Gupta} 
may be drastically enhanced in scenarios with extra dimensions \cite{Cheung:1999ja}.
This is not a new idea,
but strongly deserves a revisit \cite{Dobrich:2012nv} in the context of UV-complete theories,
that do not rely on an energy cutoff scale, such as asymptotically
safe gravity, see \cite{AS_reviews} for reviews. 
It is worth exploring the possibility  further if laser-based searches 
could eventually even shed light onto QG.

\section{Summary}

Laser-based laboratory searches can be a strong tool to address various questions of 
beyond the Standard Model physics. In this contribution we have discussed the status and prospect of
ALPS-II, which is designed to explore particularly motivated parameter regions of different WISPs.
Further, we have briefly pointed at the possibility of even testing quantum gravity scenarios with purely 
laser-based setups.

\vspace{0.1cm}
\noindent
{\it The author would like to thank the Patras conference organizers for a marvelous 
and stimulating workshop in Chicago
and all fellow colleagues from ALPS-II for fruitful and fun collaboration.}


\begin{footnotesize}

\end{footnotesize}



\begin{thebibliography}{99}



\bibitem{Jaeckel:2010ni} 
  J.~Jaeckel  {\it et al.}
  Ann.\ Rev.\ Nucl.\ Part.\ Sci.\  {\bf 60}, 405 (2010);
  A.~Ringwald,
  arXiv:1210.5081 [hep-ph].

\bibitem{Wilczek:1977pj} 
  F.~Wilczek,
  Phys.\ Rev.\ Lett.\  {\bf 40}, 279 (1978);
  S.~Weinberg,
  Phys.\ Rev.\ Lett.\  {\bf 40}, 223 (1978).
  
\bibitem{Cicoli:2012sz} 
  M.~Cicoli  {\it et al.},
  JHEP {\bf 1210}, 146 (2012)
  [arXiv:1206.0819 [hep-th]].
  
\bibitem{Arias:2012az} 
J.~Redondo, theese proceedings;
  P.~Arias {\it et al.},
  JCAP {\bf 1206}, 013 (2012).



  
  \bibitem{Hewett:2012ns} 
  J.~L.~Hewett
  {\it et al.},
  arXiv:1205.2671 [hep-ex].
  
    

  

  
\bibitem{Isern:2008nt}
    J.~Isern, these proceedings;
    J.~Isern {\it et al.},
  Astrophys.\ J.\  {\bf 682} (2008) L109
  [arXiv:0806.2807 [astro-ph]].
  
\bibitem{Horns:2012fx} 
  
    M.~Meyer  {\it et al.}, these proceedings arXiv:1211.6405;
    D.~Horns and M.~Meyer,
  JCAP {\bf 1202}, 033 (2012).

  

  
\bibitem{Abel:2008ai} 
  S.~Andreas, these proceedings
  arXiv:1211.5160;
  S.~A.~Abel {\it et al.},
  JHEP {\bf 0807}, 124 (2008).


\bibitem{Jaeckel:2008fi}
  J.~Jaeckel, J.~Redondo, A.~Ringwald,
  Phys.\ Rev.\ Lett.\  {\bf 101}, 131801 (2008).
  [arXiv:0804.4157 [astro-ph]].
  




\bibitem{Brax:2012ie} 
  C.~Burrage, these proceedings; P.~Brax {\it et al.}
  JCAP {\bf 1210}, 016 (2012)
  [arXiv:1206.1809 [hep-th]].
  
  


\bibitem{Redondo:2010dp} 
  J.~Redondo {\it et al.}
  Contemp.\ Phys.\  {\bf 52}, 211 (2011);
  P.~Arias {\it et al.}
  Phys.\ Rev.\ D {\bf 82}, 115018 (2010).
  
\bibitem{Dobrich:2012sw} 
 F.~Karbstein, these proceedings; B.~Dobrich {\it et al.},
  Phys.\ Rev.\ Lett.\  {\bf 109}, 131802 
(2012)
 \&  arXiv:1203.4986 





  
\bibitem{Ehret:2010mh}
  K.~Ehret {\it et al.},
  Phys.\ Lett.\  B {\bf 689}, 149 (2010)
  [arXiv:1004.1313 [hep-ex]].
  
\bibitem{vonSeggern:2011em} 
  J.~E.~von Seggern [ALPS Collaboration],
  ``Status of ALPS-II at DESY,'' DESY-PROC-2011-04
  




\bibitem{TDR}
Any Light Particle Search II -- Technical Design Report (2012), internal document, to be published

  
    \bibitem{baehre}
R.~B\"ahre, these proceedings; B.~Willke contribution to ``Vistas in Axion Physics'', Seattle 2012


\bibitem{Hoogeveen:1990vq} 
  F.~Hoogeveen and T.~Ziegenhagen,
  Nucl.\ Phys.\ B {\bf 358}, 3 (1991).
  

  
\bibitem{Betz:2012tp} 
M.~Betz, these  proceedings;
  M.~Betz  {\it et al.},
  Conf.\ Proc.\ C {\bf 1205201}, 3320 (2012).
 

\bibitem{Mueller:2010zzc} 
  G.~Mueller, P.~Sikivie, D.~B.~Tanner and K.~van Bibber,
  AIP Conf.\ Proc.\  {\bf 1274}, 150 (2010).
  
\bibitem{Kando:2008te} 
  M.~Kando  {\it et al.},
  AIP Conf.\ Proc.\  {\bf 1024}, 197 (2008);
  T.~Tajima  {\it et al.},
  Prog.\ Theor.\ Phys.\  {\bf 125}, 617 (2011).



  
  \bibitem{Gupta}
B.~B.~Barker, M.~S.~Bhatia and S.~N.~Gupta,
Phys. Rev. 158, 1498 (1967),
Erratum-ibid. 162, 1750.

\bibitem{Cheung:1999ja} 
  K.~-m.~Cheung,
  Phys.\ Rev.\ D {\bf 61}, 015005 (2000);
  H.~Davoudiasl,
  Phys.\ Rev.\ D {\bf 60}, 084022 (1999).



\bibitem{Dobrich:2012nv} 
  B.~Dobrich and A.~Eichhorn,
  JHEP {\bf 1206}, 156 (2012);
  A.~Eichhorn,
  arXiv:1210.1528 [hep-th].
  
  \bibitem{AS_reviews}
  M.~Niedermaier  {\it et al.},
  Living Rev.\ Rel.\  {\bf 9}, 5 (2006);
  M.~Reuter  {\it et al.},
  arXiv:1202.2274 [hep-th].
  




\end{thebibliography}
\end{document}